\documentclass[useAMS,usenatbib]{mn2e}

\usepackage{amsmath}
\usepackage{graphicx}
\usepackage{amssymb}

\title[ISW map recovery]
{Integrated Sachs-Wolfe effect map recovery from NVSS and WMAP 7yr data}
\author[Barreiro et al.]
{R. B. Barreiro$^{1}$, P. Vielva$^{1}$, A. Marcos-Caballero$^{1,2}$, E. Mart\'\i nez-Gonz\'alez${^1}$ \\
$^{1}$ Instituto de F{\'\i}sica de Cantabria (CSIC-Univ. Cantabria), Avda. de Los Castros s/n, E-39005 - Santander, Spain\\
$^{2}$ Departamento de F{\'\i}sica Moderna (Univ. Cantabria), Avda. de Los Castros s/n, E-39005 - Santander, Spain\\
\hspace{0.1cm}E-mails : barreiro@ifca.unican.es, \\}
\date{Accepted 2012 December 10. Received 2012 November 23; in original form 2012 August 11}
\pagerange{\pageref{firstpage}--\pageref{lastpage}}
\pubyear{1998}

\begin{document}
\maketitle
\label{firstpage}

\begin{abstract}

  We present a map of the Cosmic Microwave Background (CMB)
  anisotropies induced by the late Integrated Sachs Wolfe effect. The
  map is constructed by combining the information of the WMAP 7-yr CMB
  data and the NRAO VLA Sky Survey (NVSS) through a linear
  filter. This combination improves the quality of the map that would
  be obtained using information only from the Large Scale Structure
  data.  In order to apply the filter, a given cosmological model
  needs to be assumed. In particular, we consider the standard
  $\Lambda$CDM model. As a test of consistency, we show that the
  reconstructed map is in agreement with the assumed model, which is
  also favoured against a scenario where no correlation between the CMB
  and NVSS catalogue is considered.
\end{abstract}


\section{Introduction}
\label{intro}

Recent observations, as cosmic microwave background (CMB),
supernovae type Ia (SNIa) or baryon acoustic oscillations (BAOs)
agree in establishing a current accelerated expansion of the
Universe~\citep[see][for a recent review]{weinberg2012}, which, despite
some less popular interpretations, is believed to be caused by the
presence of some dark energy~\citep[see][for a review]{peebles2003}.
The nature of dark energy is one of the most puzzling issues in modern
cosmology. The
actual characteristics of this fluid are still unclear, although, up to
date, a good agreement is found between the observations and the
predictions derived from the presence of a cosmological constant with
an equation of state $p=-\rho$.

One of the classical probes of dark energy is given by a non-null
contribution to the CMB anisotropies from the late integrated Sachs-Wolf 
effect~\citep[ISW,][]{sachs1967}, under the assumption of a spatially flat
universe. The (linear) ISW fluctuations are higher at very large angular
scales, but, in any case, much smaller than the primary CMB fluctuations. In a
seminal work,~\cite{crittenden1996} proposed the cross-correlation of the
CMB fluctuations and the dark matter distribution (typically traced by
galaxy catalogues) as a possible approach to detect the ISW. Soon after
the release of the WMAP data,~\cite{boughn2004} reported the first detection
of the ISW effect via the CMB and the galaxy number density field. Posterior works
\citep[e.g.,][]{fosalba2003,vielva2006,pietrobon2006,ho2008,giannantonio2008,mcewen2008,dupe2011,schiavon2012}
have confirmed the detection of the ISW effect by exploring several galaxy
catalogues and cross-correlation techniques. Average detection is found at $\approx 3\sigma$.

Besides these works focused on the statistical detection of the ISW,
more recently there have been attempts to recover the actual ISW
fluctuations on the sky.  In principle, an optimal ISW map could be
derived from a 3D gravitational potential. This has been explored from
very large simulations~\citep[e.g.,][]{cai2010}. The complexity to
recover optimally the potential from surveys of galaxies with redshift
information is challenging~\cite[e.g.,][]{kitaura2009,jasche2010}, but
very promising from the ISW studies point of
view~\cite[e.g.,][]{frommert2008}.

Other approaches can be followed from surveys where the redshift information is
poor, or known only statistically.~\cite{barreiro2008} proposed to use jointly
maps of CMB anisotropies and of the galaxy number density field to recover
the ISW signal on the sky. Alternative works making 
use only of galaxy catalogue maps
have been proposed
afterwards: \cite{granett2009} on LRGs from SDSS-DR6,
or~\cite{francis2010a} and~\cite{dupe2011} on 2MASS.

This paper presents an application of the approach described in~\cite{barreiro2008}
to WMAP \citep{jarosik2011} and NVSS~\citep{condon1998}. The outline of the article is as follows.
The methodology is reviewed in Section~\ref{sec:method}.
A description of the data used, as well as the fiducial theoretical model is presented
in Section~\ref{sec:data}. In Section~\ref{sec:analysis} we present the results.
Finally, conclusions are given in Section~\ref{sec:final}.

\section{Methodology}
\label{sec:method}

In order to reconstruct the ISW map, we have used the linear
covariance-based filter presented in \cite{barreiro2008}. We give here the
outline of the method.

Since the filter is implemented in harmonic space, for simplicity, we
will assume that the considered data sets are full-sky. Let us denote $s_{\ell m}$
and $g_{\ell m}$ to the harmonic coefficients of the ISW map and the
large-scale structure (LSS) survey respectively. The covariance matrix $\mathbf{C}(\ell)$ of the
signals at each multipole $\ell$ is given by
\begin{equation}
\label{eq:covariance}
\mathbf{C}(\ell)
=
\left(
\begin{array}{cc}
C_{\ell}^g & C_{\ell}^{sg} \\
C_{\ell}^{sg} & C_{\ell}^{s} \\
\end{array}
\right)
\end{equation}
where $C_{\ell}^g$ and $C_{\ell}^s$ correspond to the auto spectra of
the galaxies and ISW maps~\footnote{Note that $\mathbf{C}_{\ell}^g$ corresponds
  to the power spectrum of the observed galaxies map and thus, in a
  general case, it will contain a noise contribution, while
  $C_{\ell}^s$ refers to the power spectrum of an ideal ISW
  map.}, respectively, while $C_{\ell}^{sg}$ is the cross power
between both signals. To construct the filter, we will make use of the
Cholesky decomposition of the covariance matrix, which satisfies
$\mathbf{C}(\ell)=\mathbf{L}(\ell)\mathbf{L}^T(\ell)$, where $\mathbf{L}(\ell)$ is a lower triangular
matrix. It can be trivially shown that the elements of the Cholesky
matrix relate to the elements of $\mathbf{C}(\ell)$ as $L_{11}=\sqrt{C_\ell^g},
L_{12}=C_\ell^{sg}/\sqrt{C_\ell^g}$ and
$L_{22}=\sqrt{\left|\mathbf{C}(\ell)\right|/C_\ell^g}$, where
$\left|\mathbf{C}(\ell)\right|$ is the determinant of the covariance matrix at
each $\ell$ mode.

The estimated ISW map $\hat{s}_{\ell m}$ at each harmonic mode is given by (see \citealt{barreiro2008} for details)
\begin{equation}
\label{eq:rec}
\hat{s}_{\ell m}=\frac{L_{12}(\ell)}{L_{11}(\ell)}g_{\ell m} + \frac{L_{22}^2(\ell)}{L_{22}^2(\ell)+C_{\ell}^n}\left( d_{\ell m}-\frac{L_{12}(\ell)}{L_{11}(\ell)}g_{\ell m}\right)
\end{equation}
where $d_{\ell m}$ are the harmonic coefficients of the CMB map and $C_{\ell}^n$ is the power spectrum of the CMB signal without
including the ISW.  Therefore, to reconstruct the ISW map, we need to
assume an underlying cosmological model that determines the auto and
cross power spectra present in the previous equation.

It is interesting to note that the final reconstructed map has two
contributions: the first term in the previous equation is given by a
filtered version of the galaxies map while the second expression is a
Wiener filter (WF,~\citealt{wiener1949}) of a modified CMB data map. This modified data are simply
constructed as the original CMB map minus the filtered survey.  In the
case that there is not correlation between the CMB and the LSS survey,
the filter simply defaults to the WF of the CMB map: since
there is not correlation between both signals, the galaxies map does not
contribute to the final ISW reconstruction. If the information provided by the 
CMB map were not
considered, the estimated ISW would be given just by the filtered galaxies
map.

It can be easily shown that the expected value of the power spectrum
of the estimated ISW is given by
\begin{equation}
\label{eq:cl_lcb}
\left< C_\ell^{\hat{s}} \right>= \frac{(C_\ell^{sg})^2\left(\left|\mathbf{C}(\ell)\right|+C_\ell^g
C_\ell^n\right)+\left|
\mathbf{C}(\ell)\right|^2}{C_\ell^g\left(\left|\mathbf{C}(\ell)\right|+C_\ell^g C_\ell^n\right)}
\end{equation}
It is well known that the power spectrum of the WF reconstruction is
biased towards values lower than the true signal, with the bias
depending on the signal-to-noise ratio of the data. Since our signal
is partially reconstructed using this filter, it will also be
biased. The larger the cross-correlation between CMB and the
considered galaxies catalogue, the smaller the bias, since the WF part
will contribute relatively less than the filtered survey term (see
\citealt{barreiro2008} for details).

It is also straightforward to show that the expected cross-correlation
between the recovered signal and the galaxies catalogue is equal to
that of the assumed model. However, this is only correct if the
assumed cross and auto power spectra reflect the underlying
statistical properties between the ISW and the LSS survey. For
  instance, if we assume a non-vanishing cross-correlation in our
  model, while the data have zero correlation, our reconstructed ISW
  map would actually present a non-zero correlation with the LSS
  survey. However, this spurious correlation, whose expected value can
  be easily derived from equation (\ref{eq:rec}), would be different from the one
  assumed in our model. In practice, the difference between the
  expected values of the cross-correlation for two (reasonably)
different models will in general be small and, given the weakness of
the signal and the statistical uncertainties, it may be difficult to
discriminate between them. In any case, as we will see in
Section~\ref{sec:analysis}, the comparison between the expected and
estimated values of the cross-correlation is an interesting
consistency check.

The previous description assumes that full-sky data are
available. However, in practice, a mask will be needed to exclude
those regions, in both CMB and galaxies survey maps, that have not
been observed or are too contaminated to be included in the analysis.
In \cite{barreiro2008} was shown that the method was robust against the
presence of a mask and that the quality of the reconstruction was not
significantly affected. Therefore, to deal with this problem, we will
simply substitute in the previous equations the harmonic coefficients
and power spectra by those obtained after masking the data with the
considered mask.  In particular, the {\it masked} version of the
fiducial model for the power spectra will be obtained {\it ala}
MASTER~\citep{hivon2002}. In addition, we will make use of an apodised version of the
considered mask in order to reduce the correlations between harmonic
modes that are introduced on an incomplete sky.

\section{Data description}
\label{sec:data}

\begin{figure}
\includegraphics[width=8cm,keepaspectratio]{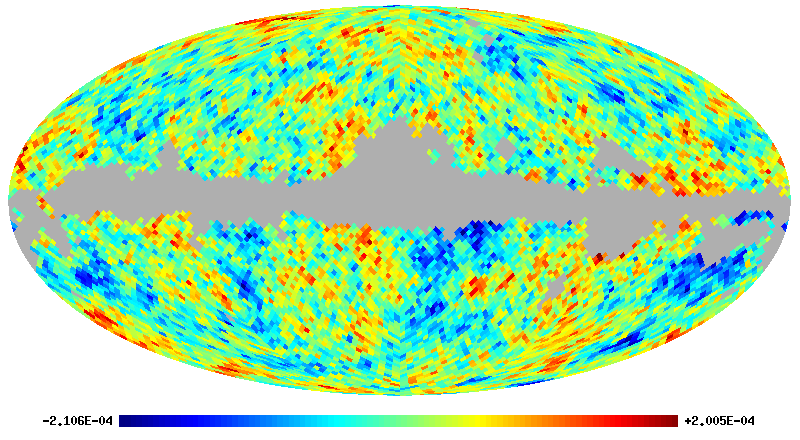}
\includegraphics[width=8cm,keepaspectratio]{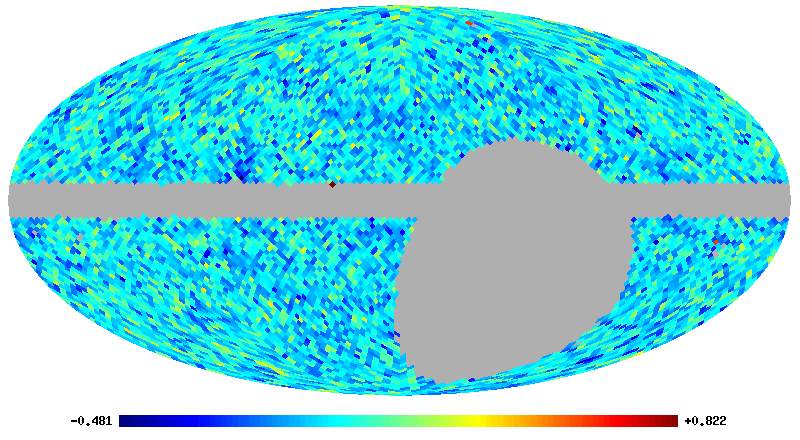}
\includegraphics[width=8cm,keepaspectratio]{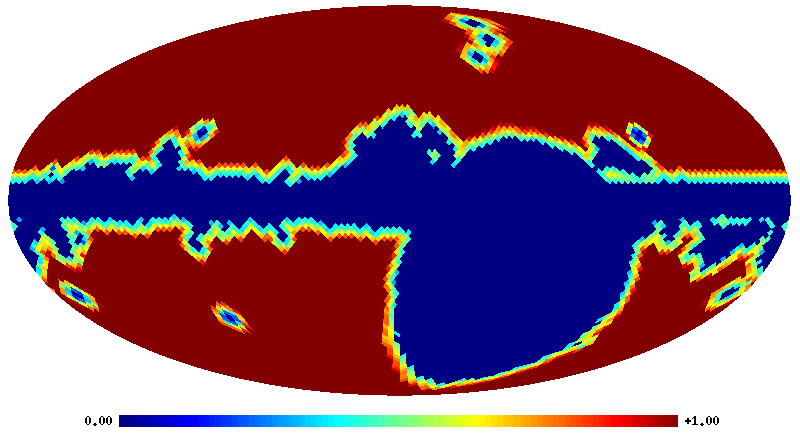}
\caption{\label{fig:data}Top: WMAP 7-yr data (in Kelvin thermodynamic temperature). Middle: NVSS galaxy density number field. Bottom:  apodised mask used for the analysis.}
\end{figure}

In order to reconstruct the ISW map using the previous methodology, we
need both a CMB map and a LSS catalogue, as well as a cosmological fiducial
model.  For the latter, we have assumed a $\Lambda$CDM model that best
fits 7-yr WMAP data, BAO and $H_0$ measurements~\citep{komatsu2011}.

For the CMB, we have made use of the 7-yr WMAP
data~\citep{jarosik2011} publicly available at the Legacy Archive for
Microwave Background Data
Analysis~\footnote{http://lambda.gsfc.nasa.gov/}. In particular, we
have constructed a CMB map as a noise-weighted combination of the foreground reduced
V- and W-band maps. The map has then been downgraded to a HEALPix
resolution of $N_{side}=32$. To reduce the galactic foreground
contamination, the KQ85 mask provided by the WMAP team has been
used. However, since we have to downgrade the mask to a lower
resolution, we have not included in the mask the holes due to point
sources since their contribution is expected to be negligible at the
considered scales. The WMAP data, with the considered mask applied, are
given in the top panel of Fig.~\ref{fig:data}.

Regarding the LSS catalogue, we have used he NRAO VLA Sky 
Survey~\citep[NVSS,][]{condon1998}. NVSS covers the north
hemisphere of the sky and part of the south until $\delta =
-40^\circ$. The NVSS catalogue has nearly $2\times10^6$ discrete
sources with fluxes above $2.5 \ \mathrm{mJy}$. The catalogue was
made using two different configurations (D and DnC) of the 
Very Large Array (VLA). This introduced a declination dependence in the number
of sources, specially for the faintest ones. Sources above
$5.0$ mJy have been selected to build a catalogue not affected by
this declination effect. The NVSS mask is defined by the unobserved sky, plus
the sources within $7$ degrees from the galactic plane, as
well as some regions associated to very near objects. The complete mask associated to NVSS leaves $73\%$
of the sky. The map is shown in the middle panel of Fig.~\ref{fig:data}.

The ISW-NVSS angular cross-power spectrum is calculated using the galaxy
redshift distribution proposed by \cite{dezotti2010}: a polynomial fit 
to the CENSORS data \citep{brookes2008}. For the galaxy bias is assumed a redshift
dependence as the one proposed by \cite{xia2011}, where the bias is just
defined by the minimum halo mass associated to NVSS radio-galaxies. The value 
of this mass is determined by fitting the NVSS angular power 
spectrum~\citep[see][, in preparation, for details]{marcos-caballero2012}.

As it is well known~\citep[e.g.][]{hernandez-monteagudo2010}, there is
a clear discrepancy between the measured NVSS auto power spectrum and
the theoretical model as described above.  In particular, an excess of
power at low multipoles is observed.  This deviation does not seem to
be related to the declination effect previously mentioned, since it is
observed even for the angular power spectrum estimated from very
bright sources. A possible explanation in terms of a non-linear
evolution of the galaxy bias (caused by a primordial non-Gaussianity
of the primordial perturbations) has been considered
by~\cite{xia2011}, although it would imply a value of the non-linear
coupling parameter $f_{\mathrm{NL}}$ larger than the one obtained from
other observables~\citep[as the CMB bi-spectrum estimated from WMAP;
  see, for instance,][]{curto2012}. Currently, the nature of this
discrepancy is not clear.  Since the filter described in the previous
section (see equation~\ref{eq:rec}) is very sensitive to the value of
the auto power spectrum of the galaxy survey, we have used for
$C_\ell^g$ a smooth fit to the observed NVSS spectrum (plus a
Poissonian term contribution) instead of the theoretical model.

Finally, we have combined the WMAP and NVSS masks, leaving a useful
area of $66\%$. The mask is then apodised with a cosine function
  (using 3 pixels, i.e. $\sim$5.5 degrees, for the size of the
  transition region with values between 0 and 1), in order to reduce
the correlations among the harmonic modes of the incomplete sky. This
apodised mask (see bottom panel of Fig.~\ref{fig:data}) is applied to
the CMB and NVSS data prior to the reconstruction of the ISW map. In
addition, the monopole and dipole outside the combined mask are
subtracted prior to the construction of the ISW map.

\section{ISW reconstruction}
\label{sec:analysis}

\begin{figure*}
\includegraphics[width=17cm,keepaspectratio]{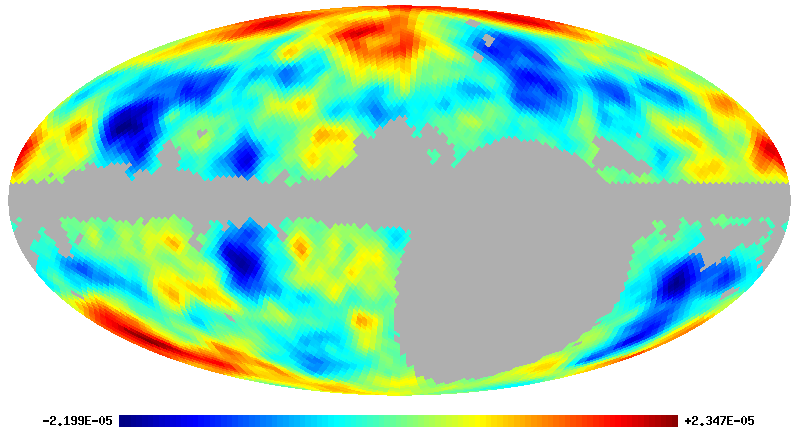}
\caption{\label{fig:iswmap} Reconstructed ISW map. Units are Kelvin (thermodynamic temperature).}
\end{figure*}
Following the described methodology, we have produced an ISW map from the WMAP and NVSS data.
The map is given in Fig.~\ref{fig:iswmap} and has been constructed at
a HEALPix resolution of $N_{side}=32$ and up to a maximum multipole $l=95$. 

\begin{figure}
\includegraphics[width=8cm,keepaspectratio]{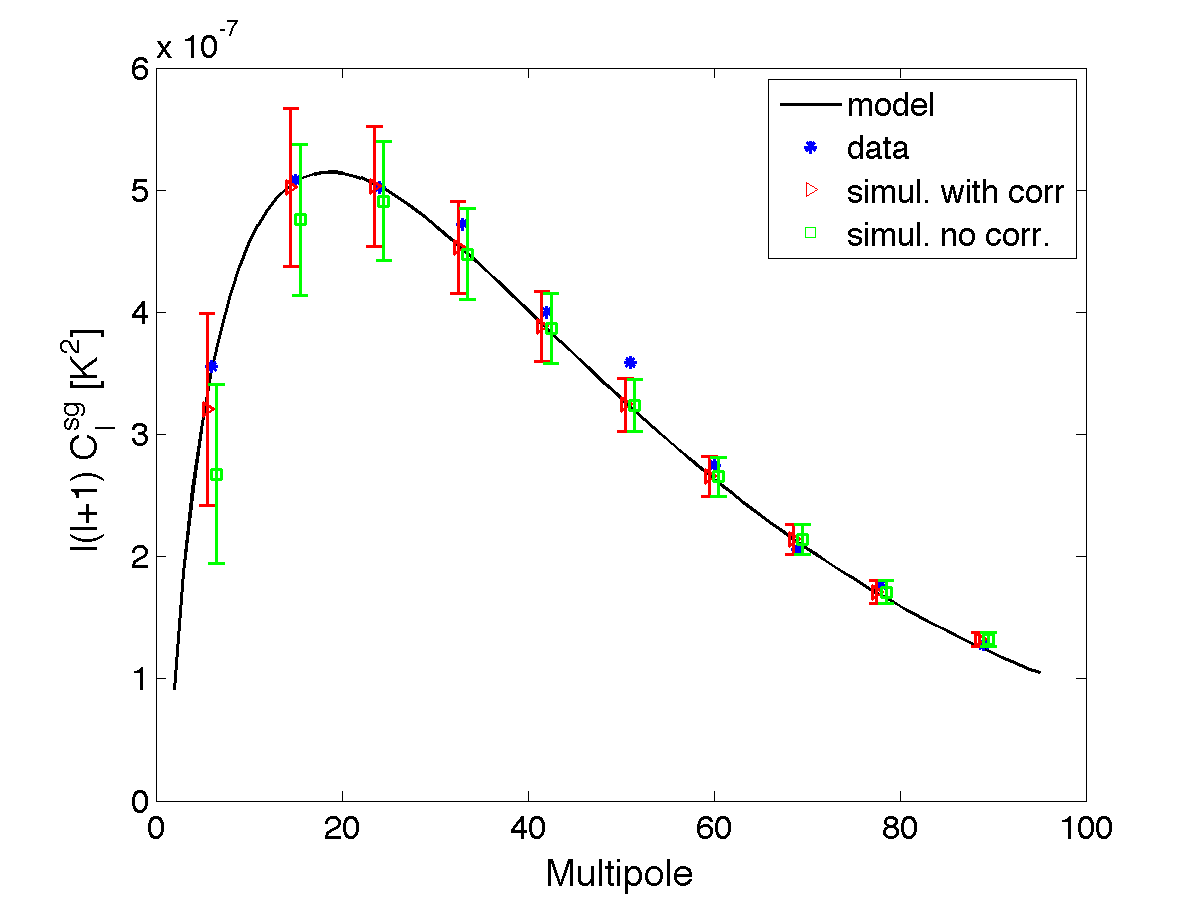}
\caption{\label{fig:cross}The binned cross-correlation between the reconstructed ISW map and
  the NVSS data is given (blue asterisks). For comparison, the
  fiducial model (black line) and the average and dispersion from
  two sets of 10000 simulations are also shown. The red triangles and their
  error bars correspond to CMB and NVSS-like simulations which are
  correlated, while for the second set of simulations (green squares
  are error bars), no correlation is introduced. The width of the bins is $\Delta_{\ell}=9$, 
  except for the last bin that is $\Delta_{\ell}=13$. For a better visualization, the results for the simulations
  have been shifted with respect to the central value of the bin.}
\end{figure}

As a consistency test, we have obtained the cross correlation between
the reconstructed ISW map and the NVSS data and compared it with the
fiducial model. The former has been simply obtained as the pseudo
cross spectrum from the masked maps, then corrected {\it ala
  MASTER} and finally binned. In addition, we have also repeated the same procedure on
two sets of 10000 simulations. For the first set, we have constructed
correlated CMB and NVSS-like simulations (generated as explained in
\citealt{barreiro2008}), assuming the same power spectra
as for reconstructing the ISW map. For the second set, the auto spectra
are the same but zero correlation is assumed between the CMB and NVSS
simulations. In both cases, we have included the corresponding
Poissonian noise in the simulations of the galaxies catalogue.
The results for this test are given in Fig.~\ref{fig:cross}.

As expected, the average value (red triangles) for the correlated
simulations agrees well with the considered model (black line). More
interestingly, the data (blue asterisks) are also compatible with
these simulations; this provides a good consistency check for our
procedure and indicates that the assumed cross and auto spectra are
reasonably close to the true statistical properties of the sky.

Conversely, the average obtained from the uncorrelated simulations
(green squares) are biased with respect to the assumed cross
spectrum. This is expected, since the assumed model is not the same as
the underlying true power spectrum.  In particular, if we assume a
  model with positive correlation for simulated data which actually have
  zero correlation, the recovered ISW map would present a spurious
  correlation with the NVSS data. More specifically, from equation
  (\ref{eq:rec}), it is easy to show that this spurious correlation is, on average, 
  lower than the one assumed for the model. Therefore, this could
  serve as a consistency test to check if the considered model describes
  well the data. Unfortunately, since the signal is very weak, the
difference between both cases is also small and it is difficult to
distinguish between both hypotheses with this test.

To try to quantify which hypothesis is favoured by the data,
we have calculated the goodness-of-fit difference: 
\begin{eqnarray}
\Delta \chi^2 & = & \chi^2_c - \chi^2_u,  \nonumber \\
\mathrm{where} & &\nonumber \\ 
\chi^2_i & = & \mathbf{c_{sg}}\mathbf{M_{i}^{-1}}\mathbf{c_{sg}^T}, 
\end{eqnarray}
$i\equiv\left\lbrace\mathrm{c, u}\right\rbrace $, and $\mathbf{c_{sg}}$ correspond to the vector constructed with the
values of the binned cross power spectrum (the binning scheme is the same as that of Fig.~\ref{fig:cross}). $\mathbf{M_{c}}$ is the
covariance matrix of the cross-spectrum obtained from 10000 correlated
simulations and $\mathbf{M_u}$ the same matrix for the uncorrelated
simulations set. We have calculated the distribution of this quantity
from independent sets of 10000 correlated and uncorrelated simulations
and compare it to the value found for the data. In particular, we find
$\Delta \chi^2 (\mathrm{data})= -2.4$. For the correlated CMB and galaxies survey
simulations, we find that 28.9 per cent
of the simulations have a value larger than the one obtained for the
data. This percentage is reduced to 6.7 per cent, when we compare the
data with the distribution obtained for uncorrelated
simulations. Therefore, although the data are in fact compatible with
the two cases considered, it favours the hypothesis of having an
underlying correlation. This further confirms the consistency of our
results.

Although, given the weakness of the considered cross-correlation, the
power of these tests is somewhat limited, the fact that the data show
consistency with the assumed model should not be taken for granted,
since other assumptions may also affect the results. For instance, if
we use the theoretical model to construct the auto spectrum of the NVSS map,
instead of a smooth fit to the measured data (as explained in
section~\ref{sec:data}), we find a clear departure of the cross
correlation between the reconstructed ISW map and the NVSS data with
respect to the expected value.

Using simulations, we have also studied which is the
improvement from combining both the CMB and NVSS data, with respect to
use only the galaxies catalogue (as for instance in~\citealt{dupe2011}
whose proposed reconstruction corresponds to the term given by the
filtered survey of equation~\ref{eq:rec}). In particular, we find an
improvement of 15 per cent in the error of the ISW reconstruction when
the CMB data are included. In any case, the relative contribution to
the reconstruction of the CMB and LSS data depends significantly on the
cross correlation between them, so it will differ for other galaxies
surveys.

\begin{figure}
\includegraphics[width=8cm,keepaspectratio]{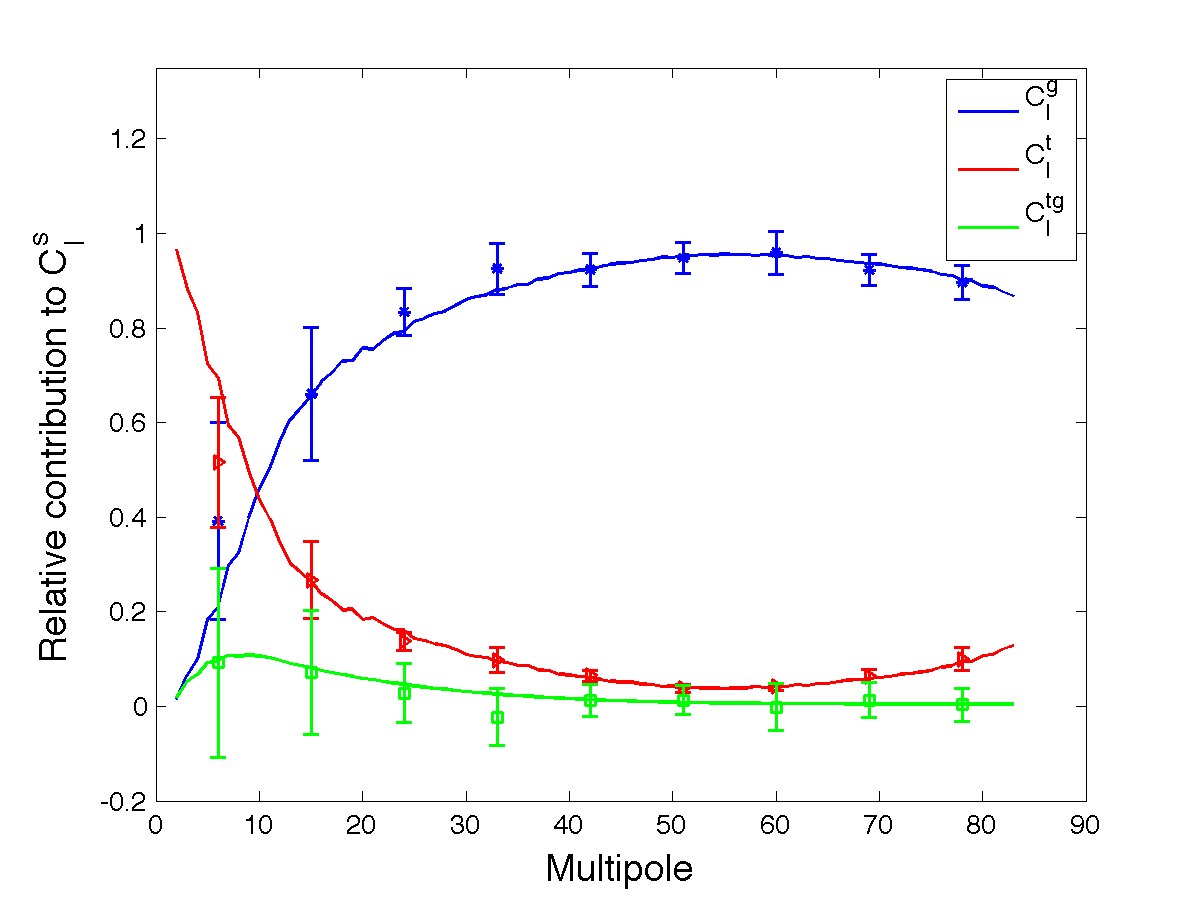}
\caption{\label{fig:relative}The expected relative contribution of the CMB and NVSS auto and cross spectra to the power spectrum of the ISW reconstruction is shown (solid lines) as well as the same quantities for the data. The results for the data have been binned to allow for a clearer comparison. The error bars correspond to the dispersion of the values within the considered bin.}
\end{figure}

In a related matter, we can also calculate the expected relative contribution of the auto and cross power of the CMB and the galaxy number density maps to the angular power spectrum of the reconstructed ISW map (using equation~\ref{eq:rec}). This is shown in Fig.~\ref{fig:relative} for the fiducial model as well as the actual contribution of each angular power spectrum for the considered data. As can be seen, the contribution from the power spectrum of the CMB $C_{\ell}^{t}$ is dominant at low multipoles (up to $\ell \approx 10$), while $C_{\ell}^{g}$ gives the main weight at higher multipoles. The contribution from the cross power between the CMB and NVSS is subdominant at all multipoles. The results obtained from the data agree quite well with the expected value. However, it is interesting to point out that the relative importance of $C_{\ell}^{t}$ with respect to $C_{\ell}^{g}$ for the data is lower than the expectation value. This can be easily explained by the fact that the amplitude of the low multipoles of the WMAP data is lower than that of the fiducial model.

\section{Conclusions}
\label{sec:final}
We have presented a reconstructed map of the ISW effect, obtained by
combining through a linear filter CMB and LSS data. In particular, we
have used the 7-yr WMAP data and the NVSS galaxies catalogue. The
joint combination of both data sets improves by a 15 per cent the error
of the ISW reconstruction in comparison to the case when only the NVSS
data are used. We have performed a consistency test, showing a good
agreement between the cross correlation inferred from the
reconstructed ISW map and the assumed fiducial model. In particular, the data
favour a $\Lambda$CDM model with respect to a scenario with null
correlation between the CMB and the NVSS data.
The relative contribution to the angular power spectrum of the ISW reconstructed
map is dominated by the CMB fluctuations up to $\ell \approx 10$ and by the
density number galaxies field at larger multipoles.

The presented methodology works in harmonic space, which implies the use of
surveys with large sky coverage, in order to avoid the
problematics introduced by large masks. However, this technique can be
easily extended to work directly in the pixel space, which would make
straightforward to deal with a mask and, in particular, LSS surveys with
smaller sky coverage could also be used to reconstruct the ISW 
signal~\citep[][in preparation]{bonavera2012}. In addition, several catalogues can be combined at the
same time, provided the covariance matrix between the surveys and
the CMB data is known.

Finally, let us remark that the application of the approach described in
this paper to future surveys (as EUCLID~\citealt{refregier2010}
or J-PAS~\citealt{benitez2009}, with very large sky coverage and very
accurate redshift estimation) could provide maps of the ISW
anisotropies caused by the large-scale structure at different
redshift shells. This will provide a tomographic view of the
ISW fluctuations.

\section*{Acknowledgements}

The authors thank Andr\'es Curto, Ra\'ul Fern\' andez-Cobos and
Francesco Paci for useful discussions.  We acknowledge partial
financial support from the Spanish Ministerio de Econom{\'\i}a y
Competitividad AYA2010-21766-C03-01 and Consolider-Ingenio 2010
CSD2010-00064 projects. PV also acknowledges financial support from
the Ram\'on y Cajal programme and AMC thanks the Spanish Ministerio de Econom{\'\i}a y
Competitividad for a predoctoral fellowship. We acknowledge
the use of Legacy Archive for Microwave Background Data Analysis
(LAMBDA). Support for it is provided by the NASA Office of Space
Science. The HEALPix package~\citep{gorski2005} was used throughout
the data analysis.

\bibliographystyle{mn2e}
\bibliography{ref_isw}

\label{lastpage}

\end{document}